\documentclass[pdflatex,sn-mathphys-num]{sn-jnl}

\usepackage{graphicx}%
\usepackage{multirow}%
\usepackage{amsmath,amssymb,amsfonts}%
\usepackage{amsthm}%
\usepackage{mathrsfs}%
\usepackage[title]{appendix}%
\usepackage{xcolor}%
\usepackage{textcomp}%
\usepackage{manyfoot}%
\usepackage{booktabs}%
\usepackage{algorithm}%
\usepackage{algorithmicx}%
\usepackage{algpseudocode}%
\usepackage{listings}%
\usepackage [latin1]{inputenc}

\newcommand{\bea}{\begin{eqnarray}}
	\newcommand{\eea}{\end{eqnarray}}
\newcommand{\bes}{\begin{subequations}}
	\newcommand{\ees}{\end{subequations}}

\newcommand{\ben}{\begin{equation}}
	\newcommand{\een}{\end{equation}}

\newcommand{\al}{\alpha}

\newcommand{\vphi}{\varphi}

\newcommand{\sech}{\mbox{sech}}
\renewcommand{\eta}{\vartheta}
\renewcommand{\xi}{\zeta}

\theoremstyle{thmstyleone}%
\theoremstyle{thmstyletwo}%

\theoremstyle{thmstylethree}%

\begin{document}

\title[Article Title]{Nondegenerate bright solitons and their interactions in the generalized coupled nonlinear Schr\"{o}dinger system}

\author*[1]{R. Ramakrishnan}\email{ramakrishnan.cnld@gmail.com}

\author[1]{Samudra Roy}\email{samudra.roy@phy.iitkgp.ac.in}

\author[2]{S. Stalin}\email{stalin.cnld@gmail.com}

\author[2]{M. Lakshmanan}\email{lakshman.cnld@gmail.com}

\affil[1]{\orgdiv{Department of Physics}, \orgname{Indian Institute of Technology Kharagpur}, \orgaddress{\city{Kharagpur}, \postcode{721302}, \state{West Bengal}, \country{India}}}

\affil[2]{\orgdiv{Department of Nonlinear Dynamics}, \orgname{Bharathidasan University}, \orgaddress{\city{Tiruchirappalli}, \postcode{620024}, \state{Tamil Nadu}, \country{India}}}


\abstract{It is known that the generalized coupled nonlinear Schrödinger (GCNLS) equations can be reduced to the basic vector nonlinear Schrödinger models through various symmetry reductions. By using such reductions, soliton solutions of several interesting types can be obtained for the GCNLS system. In this paper, we show how the non-degenerate soliton solutions can be derived using one such reduction and analyze the various special features associated with the resulting soliton solutions. We find that the obtained non-degenerate soliton solutions exhibit breathing behavior, characterized by a breathing frequency. We also show that the vector solitons emerging from the reduction undergo elastic collisions with the standard phase shift, similar to the non-degenerate solitons of other coupled nonlinear Schrödinger models. Further, they undergo interesting energy-sharing collisions when they interact with the already known bright solitons. These collision scenarios are further confirmed by an appropriate asymptotic analysis. We have also analyzed the stability of the obtained vector solitons and found that they are stable against random perturbations.  The results presented here enhance the understanding of the nature and dynamics of non-degenerate vector solitons.}



\maketitle

\section{Introduction}\label{sec1}
Integrable coupled nonlinear Schr\"{o}dinger (CNLS) equations occupy a central position in the theory of nonlinear wave propagation due to their ability to describe the evolution of interacting waves in a variety of physical situations especially in nonlinear optics, hydrodynamics, Bose-Einstein condensation and plasma systems. Among the multicomponent NLS models, the generalized coupled nonlinear Schr\"{o}dinger (GCNLS) system has received considerable attention because it incorporates four-wave mixing effects along with self-phase modulation and, cross-phase modulation within a single mathematical framework. In the present work, we consider the generalized coupled nonlinear Schr\"{o}dinger equations of the form 
\begin{eqnarray}
	i\psi_{1,z}+\psi_{1,tt}+2(a|\psi_{1}|^{2}+c|\psi_{2}|^{2}+b\psi_{1}\psi_{2}^{*}+b^{*}\psi_{1}^{*}\psi_{2})\psi_{1}=0, \nonumber \\
	i\psi_{2,z}+\psi_{2,tt}+2(a|\psi_{1}|^{2}+c|\psi_{2}|^{2}+b\psi_{1}\psi_{2}^{*}+b^{*}\psi_{1}^{*}\psi_{2})\psi_{2}=0. \label{1}
\end{eqnarray}
In Eq.(\ref{1}), $\psi_{j}$'s are complex optical field envelops, and the real constants $a,c$ appearing in Eq.(\ref{1}) describe the strength of self-phase and cross-phase modulations, and the complex constant $b$ appearing in Eq.(\ref{1}) defines the strength of four-wave mixing effect. When $b=0$, Eq. (\ref{1}) can be reduced to the Manakov system \cite{manakovpaper} for $a=c$. The GCNLS system (\ref{1}) is a completely integrable dynamical system for arbitrary choices of system parameters $a,c$ and $b$ (for $b \ne 0$), which has been proved through the Painlev\'{e} singularity structure analysis \cite{lu&pengnld} and by the Riemann-Hilbert approach \cite{dwangjmp}. This integrability nature of the above GCNLS system leads to the construction of several classes of exact localized solutions and allows to study their interaction properties. The multi-nondegenerate soliton solutions obtained by solving Eq. (\ref{1}) directly through the classical Hirota bilinearization method have been reported in Ref. \cite{rknld}. Also degenerate multi-soliton solutions of Eq. (\ref{1}) have already been derived through the standard Hirota bilinear technique and their collision dynamics reported in \cite{nvpriyacommun}. Here, we wish to note that the soliton and other nonlinear wave solutions were derived by solving the GCNLS system either through the associated eigenvalue problem or by solving the corresponding bilinear equations, without exploiting any linear connection between the GCNLS system and the reduced CNLS models \cite{telmanpre, yuanepl, telmannld}.

An interesting feature that exists in the GCNLS system is the linear transformation that maps Eq. (\ref{1}) with the  Manakov-type canonical models \cite{telmanpre, yuanepl, telmannld}. Such a correspondence provides a direct connection between two integrable coupled NLS systems and enables one to deduce various interesting kinds of nonlinear wave solutions. The connection between the GCNLS system (\ref{1}) and the well-known Manakov and Makhankov models have been clearly demonstrated in \cite{telmanpre}. As shown in the Refs. \cite{telmanpre, yuanepl, telmannld}, there exist a linear transformation connecting these two systems, which is obtained by symmetry operations,  of the following form
\begin{eqnarray}
	\psi_{1}&=&\alpha_{1}q_{1}-(\alpha_{1}^{*}b^{*}+\alpha_{2}^{*}c)q_{2},\nonumber \\
	\psi_{2}&=&\alpha_{2}q_{1}+(\alpha_{1}^{*}a+\alpha_{2}^{*}b)q_{2}. \label{2}
\end{eqnarray}
Here $\alpha_{j}$'s are complex coupling parameters, $q_{j}$, $j=1,2,$ are the solutions of the Manakov-type system. The above linear transformation connects the solutions of the GCNLS system ($\psi_{j}$'s) and the solutions of the following Manakov-type system,
\begin{eqnarray}
	iq_{1z}+q_{1tt}+2(\sigma_{1}|q_{1}|^{2}+\sigma_{2}|q_{2}|^{2})q_{1}=0,\nonumber \\
	iq_{2z}+q_{2tt}+2(\sigma_{1}|q_{1}|^{2}+\sigma_{2}|q_{2}|^{2})q_{2}=0,\label{3}
\end{eqnarray}
where $\sigma_{1}=a|\alpha_{1}|^{2}+c|\alpha_{2}|^{2}+b\alpha_{1}\alpha_{2}^{*}+b^{*}\alpha_{1}^{*}\alpha_{2}$, $\sigma_{2} = \sigma_{1}(ac-|b|^{2})$. This transformation (\ref{2}) enables one to uncover an interesting class of soliton solutions that cannot be obtained by directly solving the system (\ref{1}) through bilinearization method as given in \cite{rknld}. Further, in \cite{telmanpre}, the authors have constructed bright-bright, dark-dark, quasibreather-dark soliton solutions of the GCNLS system (\ref{1}) by mapping their respective degenerate soliton solutions of the Manakov-type system (\ref{3}) and studied their collision properties.  We wish to note here that the transformations used in \cite{telmanpre} can be obtained by fixing the parameters as $\alpha_{1}=1$ and $\alpha_{2}=0$ in Eq. (\ref{2}). We wish to note that a more general form of the linear transformation (\ref{2}) has been presented in \cite{telmannld} to describe the connection between the GCNLS system (\ref{1}) and the Manakov-type system (\ref{3}). In this work, the authors have observed vector dark solitons with unconventional oscillating background density using the Darboux transformation method and studied their interactions. However, in the present study, we consider a simple transformation of the form (\ref{2}) for our convenience. From these studies \cite{telmanpre, yuanepl, telmannld}, it is evident that one can also derive solutions of various types using the link between the GCNLS system and the Manakov-type system (\ref{3}) without solving Eq. (\ref{1}). In this paper, we will utilize this connection to achieve our main objective, which is outlined below.

In recent years, particular attention has been devoted to the study of nondegenerate vector solitons in integrable multicomponent systems. Unlike the degenerate vector solitons, which are characterized by a single wave number in all the field components \cite{manakovpaper, radhakrishnanjpa1, radhakrishnanpre, sheppardpre, radhakrishnanjpa2, vijayajayanthipra, fengjpa, ohtasiam}, nondegenerate vector solitons arise from distinct wave numbers associated with different modes. As a result, they exhibit interesting properties, including multi-hump profile structures and unconventional collision dynamics. Such states have been identified in several integrable coupled systems and have been shown to possess richer internal structures than their degenerate counterparts. This interesting class of nondegenerate bright solitons have been identified not only in the Manakov system \cite{ssprl, rkpre, ssmdpi} but also in other coupled systems \cite{sspla, rkjpa, sspre, sscsf, liuoe, gengnld, dingoe, songoe, zhangzamp, raocsf, gengnld2, qinpre1, qinpre2, dingcsf, abbagarinld}.

As we have pointed out above, using the transformation (\ref{2}), several interesting vector soliton solutions have been reported by mapping various forms of soliton solutions of the Manakov-type canonical models. However, to the best of our knowledge, obtaining the non-degenerate vector soliton solutions of the GCNLS system through the transformation (\ref{2}), and revealing the various features associated with the resulting solutions, along with a detailed investigation of the influence of the linear coupling and system parameters on their propagation and collision dynamics, have not yet been addressed in the literature. It is therefore natural to investigate whether the solutions arising from the transformation exhibit qualitatively novel states and to unravel their propagation and interaction dynamics. 

Motivated by these observations, in the present work we investigate the nondegenerate bright solitons of the GCNLS system generated through the exact linear transformation connecting it with the Manakov-type system. Starting from the known nondegenerate soliton solutions of the latter model, we construct a family of linearly superimposed nondegenerate vector solitons of the GCNLS system and examine their structural peoperties. We show that the resulting states can exhibit breathing behaviour whose characteristics are strongly influenced by the coupling parameters of the transformation. We further analyze the role of four-wave mixing effects and establish the limiting reduction to the previously known degenerate vector solitons. The interaction dynamics of the transformed nondegenerate solitons are then investigated through a detailed asymptotic analysis of the corresponding two-soliton solution. The analysis reveals distinct collision scenarios involving nondegenerate-nondegenerate, nondegenerate-degenerate and purely degenerate vector solitons. Based on the choice of the soliton parameters, both shape-preserving and energy-sharing interactions are shown to occur. These results demonstrate that the linear transformation provides a nontrivial mechanism for generating vector soliton dynamics beyond those usually encountered in the direct bilinearization of the GCNLS system. 

This paper is structured as follows. In Sec. \ref{sec2}, we will deduce the linearly superimposed nondegenerate one bright soliton solution of (\ref{1}) using the nondegenerate one bright soliton solution of the Manakov-type equation (\ref{3}) through the linear transformation (\ref{2}) and study their structural properties. In Sec. \ref{sec3}, we will analyse various collisional behaviours of the linearly superimposed solitons by deducing the two-soliton solution through the linear transformation (\ref{2}) and appropriate asymptotic analysis. We will also analyze the numerical stability of the deduced solitons against perturbations in Sec. \ref{sec4}. Finally we conclude the results in Sec. \ref{sec5}.
\section{Linearly superimposed nondegenerate soliton solution}\label{sec2} 

In this section, we deduce the basic nondegenerate soliton solution of the GCNLS system (\ref{1}) by using the nondegenerate soliton solution of the Manakov-type system (\ref{3}) using the linear transformation (\ref{2}). The explicit form of the latter solution as obtained in \cite{rkpre} is presented in Appendix A for convenience. By using the transformation (\ref{2}), we obtain the following form of the nondegenerate one soliton solution of the GCNLS system (\ref{1}). It can be written as follows,
\begin{eqnarray}
	\psi_{1}&=&\frac{2\alpha_{1}}{D_{1}}k_{1R}A_1e^{i\eta_{1I}}\cosh(\xi_{1R}+\phi_{1})-\frac{2(\alpha_{1}^{*}b^{*}+\alpha_{2}^{*}c)}{D_{2}}l_{1R}A_2e^{i\xi_{1I}}\cosh(\eta_{1R}+\phi_{2}),\nonumber \\
	\psi_{2}&=&\frac{2\alpha_{2}}{D_{1}}k_{1R}A_1e^{i\eta_{1I}}\cosh(\xi_{1R}+\phi_{1})+\frac{2(\alpha_{1}^{*}a+\alpha_{2}^{*}b)}{D_{2}}l_{1R}A_2e^{i\xi_{1I}}\cosh(\eta_{1R}+\phi_{2}), \label{4}
\end{eqnarray}
where
\begin{eqnarray}
	D_{1}&=&d_{11}\cosh(\eta_{1R}+\xi_{1R}+\phi_1+\phi_2+b_1)+(d_{11}^{*})^{-1}\cosh(\eta_{1R}-\xi_{1R}+\phi_2-\phi_1+b_2),\nonumber \\
	D_{2}&=&d_{12}\cosh(\eta_{1R}+\xi_{1R}+\phi_1+\phi_2+b_1)+(d_{12}^{*})^{-1}\cosh(\eta_{1R}-\xi_{1R}+\phi_2-\phi_1+b_2).\nonumber
\end{eqnarray}
Here 
$d_{11}=\frac{(k_{1}^{*}-l_{1}^{*})^{\frac{1}{2}}}{(k_{1}^{*}+l_{1})^{\frac{1}{2}}}$,  
$d_{12}=\frac{(k_{1}^{*}-l_{1}^{*})^{\frac{1}{2}}}{(k_{1}+l_{1}^{*})^{\frac{1}{2}}}$, 
$b_1=\frac{1}{2}\log\frac{(k_1^*-l_1^*)}{(l_1-k_1)}$, 
$b_2=\frac{1}{2}\log\frac{(k_1-l_1)(k_1^*+l_1)}{(l_1-k_1)(k_1+l_1^*)}$, 
$\phi_1=\frac{1}{2}\log\frac{(k_1-l_1)|\alpha_{1}^{(2)}|^2\sigma_2}{(k_1+l_1^*)(l_1+l_1)^2}$,
$\phi_2=\frac{1}{2}\log\frac{(l_1-k_1)|\alpha_{1}^{(1)}|^2\sigma_1}{(k_1^*+l_1)(k_1+k_1)^2}$,
$\eta_{1R}=k_{1R}(t-2k_{1I}z)$, 
$\eta_{1I}=k_{1I}t+(k_{1R}^2-k_{1I}^2)z$, 
$\xi_{1R}=l_{1R}(t-2l_{1I}z)$, 
$\xi_{1I}=l_{1I}t+(l_{1R}^2-l_{1I}^2)z$, 
$A_1=[\alpha_{1}^{(1)}/\alpha_{1}^{(1)*}\sigma_{1}]^{1/2}$, and 
$A_2=i[\alpha_{1}^{(2)}/\alpha_{1}^{(2)*}\sigma_{2}]^{1/2}$.
The structural and propagation properties of the basic nondegenerate soliton (\ref{4}) is characterized by seven arbitrary complex parameters $k_{1}$, $l_{1}$, $\alpha_{1}^{(1)}$, $\alpha_{1}^{(2)}$, $\alpha_{1}$, $\alpha_{2}$, $b$ and two real parameters $a$, $c$, whereas the nondegenerate bright soliton given in \cite{rknld} is described by only five arbitrary complex parameters $k_{1}$, $l_{1}$, $\alpha_{1}^{(1)}$, $\alpha_{1}^{(2)}$, $b$ and two real system parameters $a$, $c$.\\
\begin{figure}
	\centering
	\includegraphics[width=.35\linewidth]{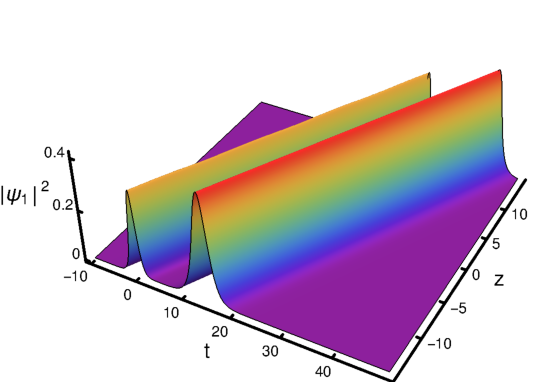}
	\includegraphics[width=.35\linewidth]{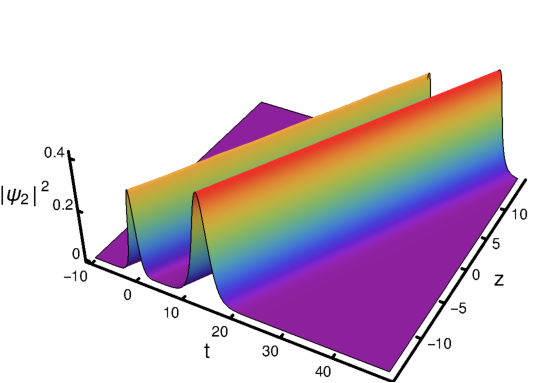}
	\caption{Double-hump nondegenerate vector solitons. Parameter values: $k_{1}=1+0.5i$, $l_{1}=0.5+0.5i$, $\alpha_{1}^{(1)}=0.65$, $\alpha_{1}^{(2)}=1+i$, $a=c=1$, $b=0.5+0.5i$ and $\alpha_{1}=\alpha_{2}=0.000001$.}
	\label{fig1}
\end{figure}
\begin{figure}
	\centering
	\includegraphics[width=.35\linewidth]{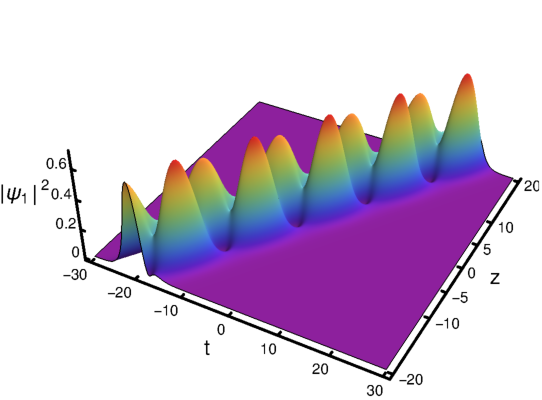}
	\includegraphics[width=.35\linewidth]{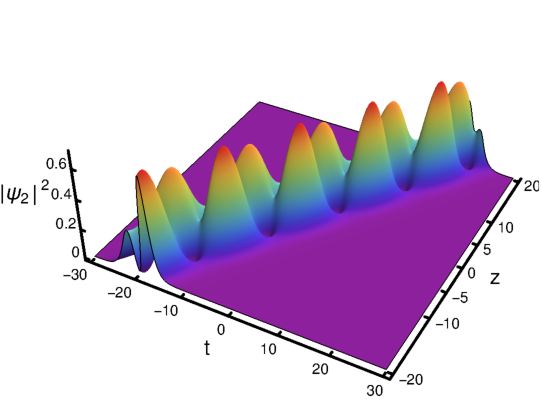}
	\caption{Double-hump breathing nondegenerate vector solitons. Parameter values: $k_{1}=1+0.5i$, $l_{1}=0.5+0.5i$, $\alpha_{1}^{(1)}=0.65$, $\alpha_{1}^{(2)}=1+i$, $a=c=1$, $b=0.5+0.5i$ and $\alpha_{1}=\alpha_{2}=1$.}
	\label{fig2}
\end{figure}
In the above solution (\ref{4}), the four-wave mixing parameter `$b$' appears in product with the linear coupling parameters $\alpha_{j}$, $j=1,2$. Therefore, the role of `$b$' is highly dependent on the strength of the linear coupling parameters $\alpha_{j}$'s. To understand this behavior, we consider two different sets of parameters $\alpha_{j}$ with other parameter values fixed as $k_{1}=1+0.5i$, $l_{1}=0.5+0.5i$, $\alpha_{1}^{(1)}=0.65$, $\alpha_{1}^{(2)}=1+i$, $a=c=1$, $b=0.5+0.5i$. In the weak linear coupling, $\alpha_{1}=\alpha_{2}=0.000001$, among the Manakov-type modes one can observe that no breathing formation occurs but simply the double-hump structure appears in both the modes which can be identified in Fig. \ref{fig1}. From Fig. \ref{fig2}, one can clearly witness the formation of breathing pattern in both the modes when we consider the strong linear coupling values, for example $\alpha_{1}=\alpha_{2}=1$, among the Manakov-type modes. This explicitly indicates that the breathing nature of the GCNLS nondegenerate bright solitons (\ref{4}) essentially relies on the linear coupling parameters $\alpha_{j}$. On the other hand, as we have shown in \cite{rknld}, the breathing nature naturally appears in nondegenerate bright solitons of the GCNLS system (\ref{1}) and depends on the strength of the four-wave mixing parameter `$b$'. We find that the linearly superimposed nondegenerate soliton solution (\ref{4}) exhibits a regular breathing nature. Such a breathing nature is demonstrated in Fig. \ref{fig2}, for the parameter values $k_{1}=1+0.5i$, $l_{1}=0.5+0.5i$, $\alpha_{1}^{(1)}=0.65$, $\alpha_{1}^{(2)}=1+i$, $a=c=1$, $b=0.5+0.5i$ and $\alpha_{1}=\alpha_{2}=1$. A regular breathing state can be observed when $ac>0$ and $ac-|b|^{2}>0$. In both the modes, this breathing pattern emerged along the $z$-direction. The period of oscillation  $T=\frac{2\pi}{|k_{1R}^{2}-l_{1R}^{2}|}$ characterises the breathing nature of the nondegenerate solitons under equal velocity conditions. It is interesting to note here that in \cite{sspla2} a new class of nondegenerate beating solitons has been generated through an entirely different kind of superposition of the Manakov nondegenerate solitons.\\
\begin{figure}
	\centering
	\includegraphics[width=.35\linewidth]{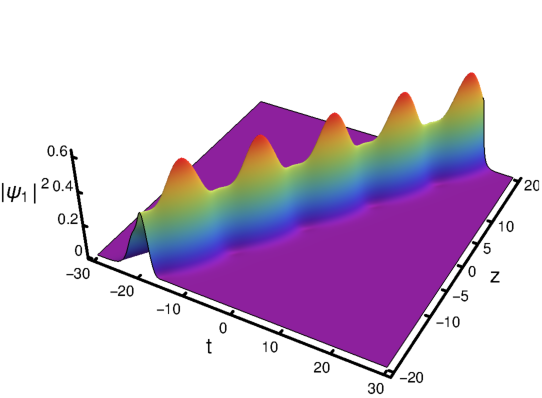}
	\includegraphics[width=.35\linewidth]{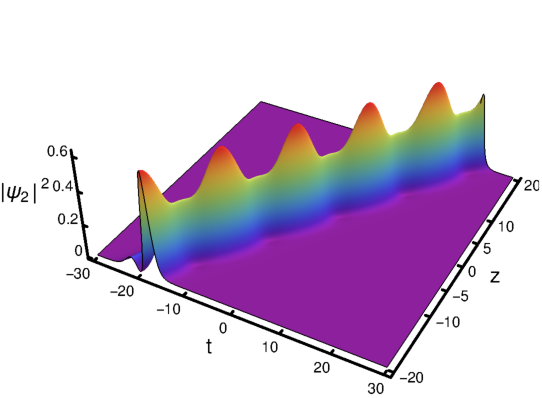}
	\caption{Double-hump breathing nondegenerate vector solitons even at $b=0$, while the values of the rest of the parameters are same as Fig. \ref{fig1}.}
	\label{fig3}
\end{figure}
\begin{figure}
	\centering
	\includegraphics[width=.35\linewidth]{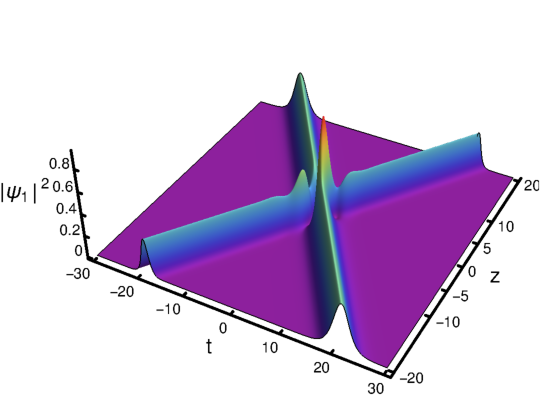}
	\includegraphics[width=.35\linewidth]{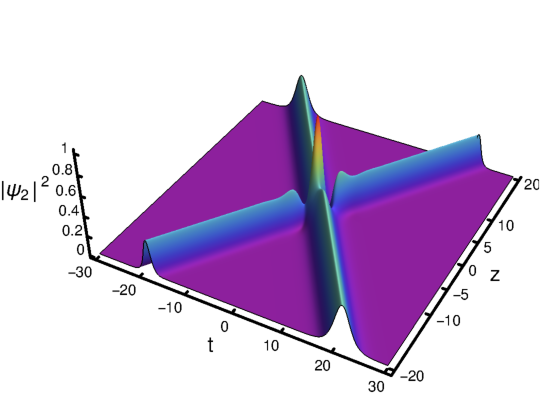}
	\caption{Nondegenerate vector soliton solution and their collision-like structures. Parameter values: $k_{1}=1+0.5i$, $l_{1}=0.5-0.5i$, $\alpha_{1}^{(1)}=0.65$, $\alpha_{1}^{(2)}=1+i$, $a=c=1$, $b=0.5+0.5i$ and $\alpha_{1}=\alpha_{2}=0.5$.}
	\label{fig4}
\end{figure}
\begin{figure}
	\centering
	\includegraphics[width=.35\linewidth]{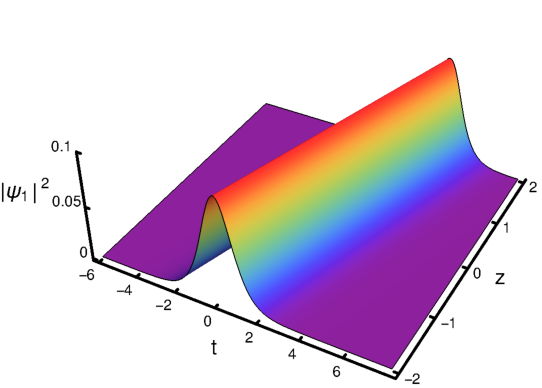}
	\includegraphics[width=.35\linewidth]{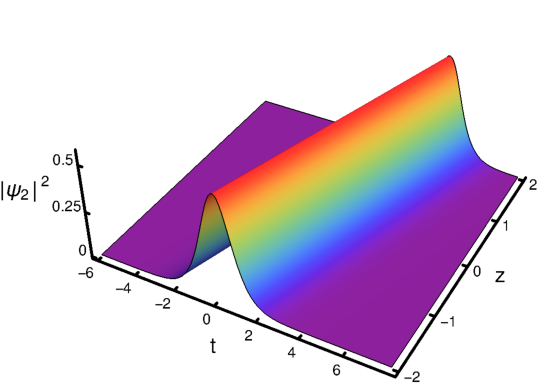}
	\caption{Purely degenerate single-hump vector solitons. Parameter values: $k_{1}=l_{1}=1+0.5i$, $\alpha_{1}^{(1)}=1+i$, $\alpha_{1}^{(2)}=0.5i$, $a=c=1$, $b=0.5+0.5i$ and $\alpha_{1}=\alpha_{2}=0.1$.}
	\label{fig5}
\end{figure}
Further, one can also observe a singular behaviour of the solution (\ref{4}) in both the defocussing nonlinear regime ($a,c<0$)  and mixed focusing-defocusing nonlinear regimes ($a>0,c<0$ or $a<0,c>0$). Interestingly we find that the solution (\ref{4}) can admit breathing behaviour even when the four wave mixing value $b=0$, as it is depicted in Fig. \ref{fig3}. Then the linearly superimposed nondegenerate soliton solution (\ref{4}) derived through the linear transformation exhibits an entirely different scenario when the relative velocity $\Delta v = v_{1}-v_{2} = 2(k_{1I}-l_{1I})= 2 \ne 0$. This can be achieved by tuning the  imaginary part of the wave numbers in the coupled modes. To demonstrate this, we fix the parameters as $k_{1}=1+0.5i$, $l_{1}=0.5-0.5i$, $\alpha_{1}^{(1)}=0.65$, $\alpha_{1}^{(2)}=1+i$, and $\alpha_{1}=\alpha_{2}=0.5$.  When the imaginary parts of the wave numbers are unequal, the velocities are different, and the soliton solution (\ref{4}) exhibits collision-like structure as given in Fig. \ref{fig4}. However, when we fix $k_{1}=l_{1}$ in (\ref{4}), the resulting solution is the purely degenerate one soliton solution of the system (\ref{1}). Under the above particular restriction on the wave numbers, $q_{1}$ and $q_{2}$ in solution (\ref{4}) reduces to the following degenerate forms, 
$q_{j}=k_{1R}B_{j}e^{i\eta_{1I}}\sech(\eta_{1R}+R/2),~ j=1,2,$
where $B_{j}=\frac{\alpha_{1}^{(j)}}{(|\alpha_{1}^{(1)}|^{2}\sigma_{1}+|\alpha_{1}^{(2)}|^{2}\sigma_{2})^{1/2}}$, $e^{R}=\frac{|\alpha_{1}^{(1)}|^{2}\sigma_{1}+|\alpha_{1}^{(2)}|^{2}\sigma_{2}}{(2k_{1R})^{2}}$, $\eta_{1R}=k_{1R}(t-2k_{1I}z)$, and $\eta_{1I}=k_{1I}t+(k_{1R}^2-k_{1I}^2)z$.
In contrast to the nondegenerate soliton, the degenerate vector bright soliton always exhibits single-hump profile structure as we shown in Fig. \ref{fig5} with the parametric choices $k_{1}=l_{1}=1+0.5i$, $\alpha_{1}^{(1)}=1+i$, $\alpha_{1}^{(2)}=0.5i$, $a=c=1$, $b=0.5+0.5i$ and $\alpha_{1}=\alpha_{2}=0.1$. In this case, the value of $k_{1R}$ determines the amplitude of the solitons while $k_{1I}$ controls the soliton velocity in both the components. Further, by carefully tuning the real part of the wave number $k_{1}$, one can easily adjust the amplitude and width of the degenerate soliton. Here, the complex parameter `$b$' behaves merely like an amplitude parameter. However, the linear transformation parameters $\alpha_{1}$ and $\alpha_{2}$ change only the central position of the solitons in this degenerate case. It is important to point out that the degenerate solitons do not admit any singularity condition. It is necessary to mention here that along with the condition $k_{1}=l_{1}$, when we fix $\alpha_{1}=1$ and $\alpha_{2}=0$ in (\ref{4}), we can exactly retain the bright-bright soliton solutions and their properties reported in \cite{telmanpre}.

\section{Collision dynamics of linearly superimposed nondegenerate vector solitons} \label{sec3}
Naturally it is of interest to investigate how the linearly transformed nondegenerate solitons of the GCNLS system (\ref{1}) interact both among themselves and with other types of solitons, such as the degenerate solitons. Another important question arises is how the breathing behaviour of the nondegenerate solitons is affected during these interactions. In particular, one may ask whether there exists any possibility of energy sharing behaviour when they interact with other types of solitons. To address these questions, we consider three distinct types of collision dynamics: (i) Interaction between nondegenerate solitons, (ii) Interaction between breathing nondegenerate soliton and a degenerate soliton, (iii) Collision between the degenerate solitons. In order to study these aspects, we consider the nondegenerate two-soliton solution of the GCNLS system (\ref{1}). This can be done by mapping the nondegenerate two-soliton solution of the Manakov-type system (\ref{3}) to the present system (\ref{1}). We have given the explicit form of the two-soliton solution of the Manakov-type equation as deduced in ref.\cite{rkpre} in Appendix A for this purpose. Using the so deduced exact form of the nondegenerate two-soliton solution of (\ref{1}) we analyse the analytical form of the individual solitons before and after the collision by carrying out an asymptotic analysis. While performing the asymptotic analysis, we restrict ourselves to the equal velocity case $k_{1I}=l_{1I}, k_{2I}=l_{2I}$. A similar analysis can be carried out for the unequal velocity case also but for brevity we omit the associated details.

\subsection{Asymptotic analysis}

To begin with, we analyse the collision dynamics among two breathing nondegenerate solitons, derived through the linear transformations (\ref{2}). For this purpose, we perform the asymptotic analysis on the linearly transformed nondegenerate two-soliton solution. The asymptotic forms of the nondegenerate solitons analysed in the limit $z \rightarrow \pm \infty$. To carry out the analysis, we consider the parametric choice as $k_{1R},k_{2R},l_{1R},l_{2R} > 0$, $k_{1I}>k_{2I}$, $l_{1I}>l_{2I}$, but $k_{1I}=l_{1I}>0$ and $k_{2I}=l_{2I}<0$ which correspond to the head-on collision between the two breathing nondegenerate solitons. Here, the wave variables are $\eta_{jR}=k_{jR}(t-2k_{jI}z)$ and $\xi_{jR}=l_{jR}(t-2l_{jI}z)$, $j=1,2$. The asymptotic behaviour of the wave variables $\eta_{jR}$ and $\xi_{jR}$ are obtained as (i) soliton 1 ($S_{1}$): $\eta_{1R}, \xi_{1R} \simeq 0, \eta_{2R}, \xi_{2R} \rightarrow \mp \infty$ as $z \rightarrow \mp \infty$ and (ii) soliton 2 ($S_{2}$): $\eta_{2R}, \xi_{2R} \simeq 0, \eta_{1R}, \xi_{1R} \rightarrow \mp \infty$ as $z \rightarrow \pm \infty$. As a result, the individual nondegenerate solitons in both the components takes the following forms.\\
\underline{(a) Soliton 1}: $z\rightarrow \mp \infty$\\
\begin{eqnarray}
	\psi_{1}^{\mp}&=&\alpha_{1}q_{1}^{1\mp}-(\alpha_{1}^{*}b^{*}+\alpha_{2}^{*}c)q_{2}^{1\mp},\nonumber \\
	\psi_{2}^{\mp}&=&\alpha_{2}q_{1}^{1\mp}+(\alpha_{1}^{*}a+\alpha_{2}^{*}b)q_{2}^{1\mp}. \label{5}
\end{eqnarray}
where
\begin{eqnarray}
	q_{1}^{1\mp}&=&\frac{1}{P_{1}^{\mp}}2k_{1R}e^{i\eta_{1I}+\theta_{1}^{\mp}}\cosh(\xi_{1R}+\phi_1^{\mp}),~
	q_{2}^{1\mp}=\frac{1}{P_{2}^{\mp}}2l_{1R}e^{i\xi_{1I}\theta_{2}^{\mp}}\cosh(\eta_{1R}+\phi_2^{\mp}),\nonumber \\
	P_{1}^{\mp}&=&c_{11}\cosh(\eta_{1R}+\xi_{1R}+\phi_1^{\mp}+\phi_2^{\mp}+d_1)+(c_{11}^*)^{-1}\cosh(\eta_{1R}-\xi_{1R}+\phi_2^{\mp}-\phi_1^{\mp}+d_2),\nonumber \\
	P_{2}^{\mp}&=&c_{12}\cosh(\eta_{1R}+\xi_{1R}+\phi_1^{\mp}+\phi_2^{\mp}+d_1)+(c_{12}^*)^{-1}\cosh(\eta_{1R}-\xi_{1R}+\phi_2^{\mp}-\phi_1^{\mp}+d_2).\nonumber
\end{eqnarray}
In the above, 
$\phi_1^-=\frac{1}{2}\log\frac{(k_1-l_1)|\alpha_1^{(2)}|^2\sigma_2}{(k_1+l_1^*)(l_1+l_1^*)^2}$,  
$\phi_2^-=\frac{1}{2}\log\frac{(l_1-k_1)|\alpha_1^{(1)}|^2\sigma_1}{(k_1^*+l_1)(k_1+k_1^*)^2}$,
$\phi_1^+=\phi_1^-+\frac{1}{2} \log\frac{|k_2-l_1|^2|l_1-l_2|^4}{|k_2+l_1^*|^2|l_1+l_2^*|^4}$, 
$\phi_2^+=\phi_2^-+\frac{1}{2}\log\frac{|k_1-l_2|^2|k_1-k_2|^4}{|k_1+l_2^*|^2|k_1+k_2^*|^4}$,
$e^{i\theta_1^-}=\frac{(k_{1}-k_{2})(l_{1}-l_{2})(l_{1}^*+l_{2})(k_{2}-l_{1})^{\frac{1}{2}}(k_{1}+k_{2}^{*})(k_{2}^{*}+l_{1})^{\frac{1}{2}}}{(k_{1}^{*}-k_{2}^{*})(l_{1}+l_{2}^*)(l_{1}^*-l_{2}^*)(k_{2}^{*}-l_{1}^{*})^{\frac{1}{2}}(k_{1}^{*}+k_{2})(k_{2}+l_{1}^{*})^{\frac{1}{2}}}$,
$c_{11}={\frac{(k_{1}^{*}-l_{1}^{*})^{\frac{1}{2}}}{(k_{1}^{*}+l_{1})^{\frac{1}{2}}}}$,
$e^{i\theta_2^-}=\frac{(l_{1}-l_{2})(k_{1}-l_{2})^{\frac{1}{2}}(k_{1}+l_{2}^{*})^{\frac{1}{2}}(l_{1}+l_{2}^{*})}{(k_{1}^{*}-l_{2}^{*})^{\frac{1}{2}}(l_{1}^{*}-l_{2}^{*})(k_{1}^{*}+l_{2})^{\frac{1}{2}}(l_{1}^{*}+l_{2})}$,
$e^{i\theta_1^+}=\frac{(k_{1}-k_{2})(k_{1}-l_{2})^{\frac{1}{2}}(k_{1}^*+k_{2})(k_{1}^{*}+l_{2})^{\frac{1}{2}}}{(k_{1}^{*}-k_{2}^{*})(k_{1}^{*}-l_{2}^{*})^{\frac{1}{2}}(k_{1}+k_{2}^*)(k_{1}+l_{2}^{*})^{\frac{1}{2}}}$,
$c_{12}=\frac{(k_{1}^{*}-l_{1}^{*})^{\frac{1}{2}}}{(k_{1}+l_{1}^{*})^{\frac{1}{2}}}$.
$e^{i\theta_2^+}=\frac{(l_{1}-l_{2})(k_{2}-l_{1})^{\frac{1}{2}}(k_{2}+l_{1}^{*})^{\frac{1}{2}}(l_{1}^*+l_{2})}{(k_{2}^{*}-l_{1}^{*})^{\frac{1}{2}}(l_{1}^{*}-l_{2}^{*})(k_{2}^{*}+l_{1})^{\frac{1}{2}}(l_{1}+l_{2}^*)}$, 
$d_1=\frac{1}{2}\log\frac{(k_1^*-l_1^*)}{(l_1-k_1)}$, 
$d_2=\frac{1}{2}\log\frac{(k_1-l_1)(k_1^*+l_1)}{(l_1-k_1)(k_1+l_1^*)}$.\\
\underline{(b) Soliton 2}: $z\rightarrow \mp \infty$\\
\begin{eqnarray}
	\psi_{1}^{\mp}&=&\alpha_{1}q_{1}^{2\mp}-(\alpha_{1}^{*}b^{*}+\alpha_{2}^{*}c)q_{2}^{2\mp},\nonumber \\
	\psi_{2}^{\mp}&=&\alpha_{2}q_{1}^{2\mp}+(\alpha_{1}^{*}a+\alpha_{2}^{*}b)q_{2}^{2\mp}. \label{6}
\end{eqnarray}
where
\begin{eqnarray}
	q_{1}^{1\mp}&=&\frac{1}{R_{1}^{\mp}}2k_{2R}e^{i(\eta_{2I}+\theta_1^{\mp})}\cosh(\xi_{2R}+\varphi_1^{\mp}),~
	q_{2}^{1\mp}=\frac{1}{R_{2}^{\mp}}2l_{2R}e^{i(\xi_{2I}+\theta_2^{\mp})}\cosh(\eta_{2R}+\varphi_2^{\mp}),\nonumber \\
	R_{1}^{\mp}&=&c_{21}\cosh(\eta_{2R}+\xi_{2R}+\varphi_1^{\mp}+\varphi_2^{\mp}+d_3)+(c_{21}^*)^{-1}\cosh(\eta_{2R}-\xi_{2R}+\varphi_2^{\mp}-\varphi_1^{\mp}+d_4),\nonumber \\
	R_{2}^{\mp}&=&c_{22}\cosh(\eta_{2R}+\xi_{2R}+\varphi_1^{\mp}+\varphi_2^{\mp}+d_3)+(c_{22}^*)^{-1}\cosh(\eta_{2R}-\xi_{2R}+\varphi_2^{\mp}-\varphi_1^{\mp}+d_4).\nonumber
\end{eqnarray}
In the above,
$\vphi_1^-=\frac{1}{2}\log\frac{(k_2-l_2)|\alpha_{2}^{(2)}|^2\sigma_2}{(k_2+l_2^*)(l_2+l_2^*)^2}+\frac{1}{2}\log\frac{|k_1-l_2|^2|l_1-l_2|^4}{|k_1+l_2^*|^2|l_1+l_2^*|^4}$,
$\vphi_2^-=\frac{1}{2}\log\frac{(l_2-k_2)|\alpha_{2}^{(1)}|^2\sigma_1}{(k_2^*+l_2)(k_2+k_2^*)^2}+\frac{1}{2}\log\frac{|k_2-l_1|^2|k_1-k_2|^4}{|k_2+l_1^*|^2|k_1+k_2^*|^4}$,
$\varphi_1^+=\frac{1}{2}\log\frac{(k_2-l_2)|\alpha_{2}^{(2)}|^2\sigma_2}{(k_2+l_2^*)(l_2+l_2^*)^2}$,
$\varphi_2^+=\frac{1}{2}\log\frac{(l_2-k_2)|\alpha_{2}^{(1)}|^2\sigma_1}{(k_2^*+l_2)(k_2+k_2^*)^2}$, 
$\varphi_3^+=\frac{1}{2}\log\frac{|k_2-l_2|^2|\alpha_{2}^{(1)}|^2|\alpha_{2}^{(2)}|^2\sigma_1\sigma_2}{|k_2+l_2^*|^2(k_2+k_2^*)^2(l_2+l_2^*)^2}$, 
$\varphi_4^+=\frac{1}{2}\log\frac{|\alpha_{2}^{(1)}|^2(l_2+l_2^*)^2\sigma_1}{|\alpha_{2}^{(2)}|^2(k_2+k_2^*)^2\sigma_2}$,
$e^{i\theta_1^-}=\frac{(k_{1}-k_{2})(l_{1}-l_{2})(l_{1}^*+l_{2})(k_{2}-l_{1})^{\frac{1}{2}}(k_{1}+k_{2}^{*})(k_{2}^{*}+l_{1})^{\frac{1}{2}}}{(k_{1}^{*}-k_{2}^{*})(l_{1}+l_{2}^*)(l_{1}^*-l_{2}^*)(k_{2}^{*}-l_{1}^{*})^{\frac{1}{2}}(k_{1}^{*}+k_{2})(k_{2}+l_{1}^{*})^{\frac{1}{2}}}$,
$e^{i\theta_2^-}=\frac{(l_{1}-l_{2})(k_{1}-l_{2})^{\frac{1}{2}}(k_{1}+l_{2}^{*})^{\frac{1}{2}}(l_{1}+l_{2}^{*})}{(k_{1}^{*}-l_{2}^{*})^{\frac{1}{2}}(l_{1}^{*}-l_{2}^{*})(k_{1}^{*}+l_{2})^{\frac{1}{2}}(l_{1}^{*}+l_{2})}$,
$e^{i\theta_1^+}=\frac{(k_{1}-k_{2})(k_{1}-l_{2})^{\frac{1}{2}}(k_{1}^*+k_{2})(k_{1}^{*}+l_{2})^{\frac{1}{2}}}{(k_{1}^{*}-k_{2}^{*})(k_{1}^{*}-l_{2}^{*})^{\frac{1}{2}}(k_{1}+k_{2}^*)(k_{1}+l_{2}^{*})^{\frac{1}{2}}}$, 
$e^{i\theta_2^+}=\frac{(l_{1}-l_{2})(k_{2}-l_{1})^{\frac{1}{2}}(k_{2}+l_{1}^{*})^{\frac{1}{2}}(l_{1}^*+l_{2})}{(k_{2}^{*}-l_{1}^{*})^{\frac{1}{2}}(l_{1}^{*}-l_{2}^{*})(k_{2}^{*}+l_{1})^{\frac{1}{2}}(l_{1}+l_{2}^*)}$,
$c_{21}=\frac{(k_{2}^{*}-l_{2}^{*})^{\frac{1}{2}}}{(k_{2}^{*}+l_{2})^{\frac{1}{2}}}$,
$c_{22}=\frac{(k_{2}^{*}-l_{2}^{*})^{\frac{1}{2}}}{(k_{2}+l_{2}^*)^{\frac{1}{2}}}$,
$d_3=\frac{1}{2}\log\frac{(k_2^*-l_2^*)}{(l_2-k_2)}$, 
$d_4=\frac{1}{2}\log\frac{(k_2-l_2)(k_2^*+l_2)}{(l_2-k_2)(k_2+l_2^*)}$.
\begin{figure}
	\centering
	\includegraphics[width=.70\linewidth]{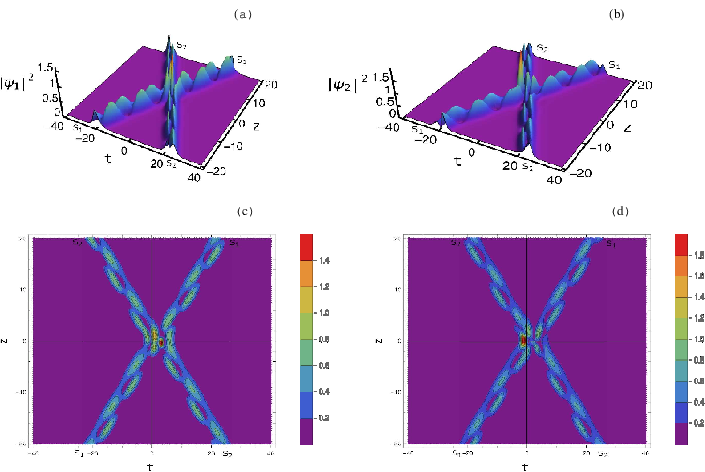}
	\caption{(a) and (b): Collision between two nondegenerate breathing solitons. (c) and (d): corresponding contour plots. The parameter values: $k_{1}=1+0.5i, l_{1}=0.5+0.5i, k_{2}=1-0.5i, l_{2}=0.5-0.5i, \alpha_{1}^{(1)}=0.65+0.5i, \alpha_{1}^{(2)}=1+i, \alpha_{2}^{(1)}=1+i, \alpha_{2}^{(2)}=0.65+0.5i, \alpha_{1}=\alpha_{2}=1, a=c=1$ and $b=0.5+0.5i$}
	\label{fig6}
\end{figure}
\begin{figure}
	\centering
	\includegraphics[width=.35\linewidth]{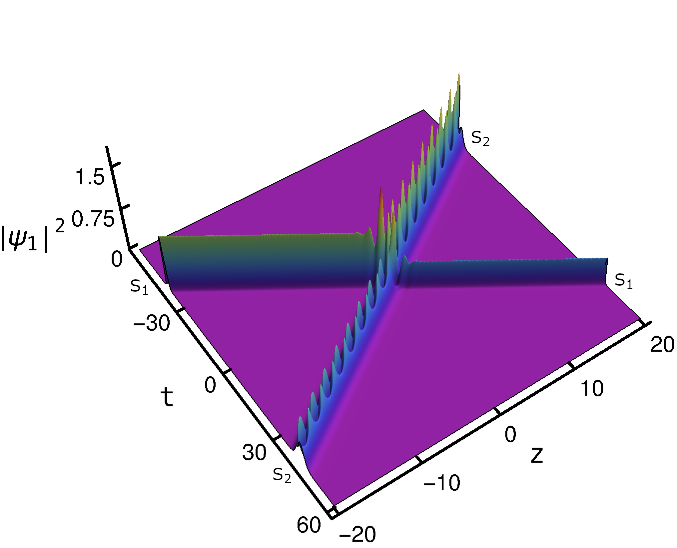}~
	\includegraphics[width=.35\linewidth]{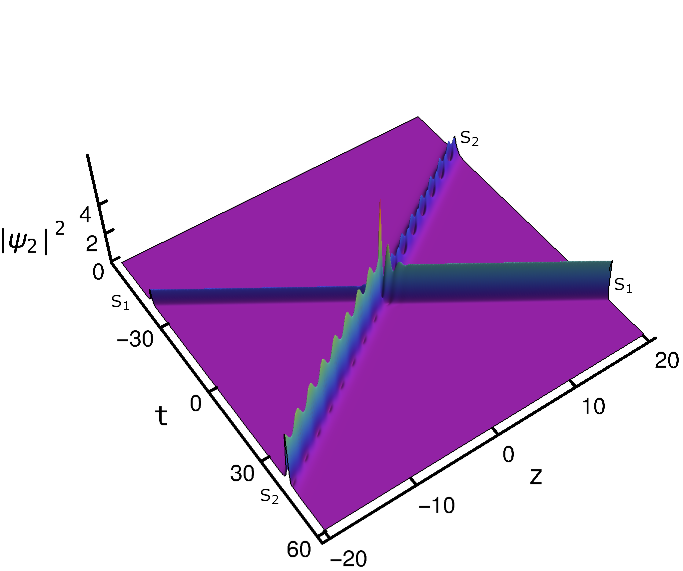}
	\caption{Collision between degenerate and nondegenerate breathing solitons. The corresponding parameter values: $k_{1}=l_{1}=1.5+i, k_{2}=1.65-i, l_{2}=0.5-i, \alpha_{1}^{(1)}=\alpha_{2}^{(1)}=0.65, \alpha_{1}^{(2)}=\alpha_{2}^{(2)}=0.5i, \alpha_{1}=\alpha_{2}=1, a=c=1$ and $b=0.5+0.5i$.}
	\label{fig7}
\end{figure}
\begin{figure}
	\centering
	\includegraphics[width=.35\linewidth]{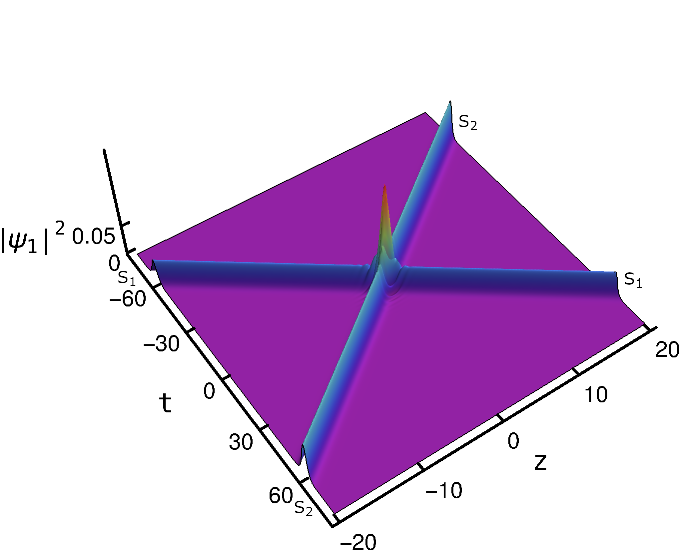}~
	\includegraphics[width=.35\linewidth]{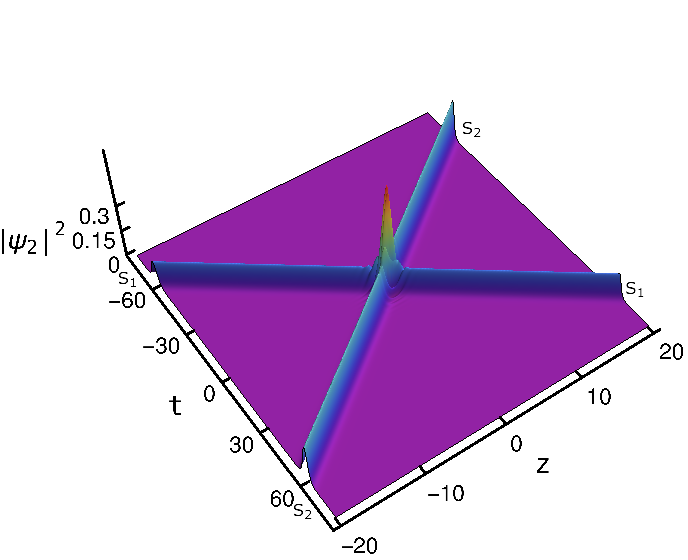}
	\caption{Shape preserving collision between degenerate single-hump solitons. The parameter values: $k_{1}=l_{1}=0.35+1.5i, k_{2}=l_{2}=0.45-1.5i, \alpha_{1}^{(1)}=\alpha_{2}^{(1)}=0.5+0.5i, \alpha_{1}^{(2)}=\alpha_{2}^{(2)}=0.45, \alpha_{1}=\alpha_{2}=1, a=c=1$ and $b=0.5+0.5i$}
	\label{fig8}
\end{figure}
In the above asymptotic expressions, superscripts $(\mp)$ indicate the corresponding solitons before and after the collision and the subscripts (1,2) represent the modes of the nondegenerate solitons.

The above asymptotic analysis explicitly shows that in general nondegenerate solitons undergo shape changing collision since the phase terms of the individual nondegenerate solitons before and after the collision are different. Also the shape preserving collision can be possible whenever the phase terms of the nondegenerate solitons obey the following conditions,
\begin{equation}
	\phi_j^-=\phi_j^+, ~ \varphi_j^-=\varphi_j^+,~ j=1,2,3,4. 
	\label{Eqphase}
\end{equation}
Indirectly the above condition demands that the nondegenerate solitons undergo the collision process without energy redistribution among the modes whenever the above condition holds.

To demonstrate this shape preserving collision between the nondegenerate solitons we fix the two-soliton parameters as $k_{1}=1+0.5i, l_{1}=0.5+0.5i, k_{2}=1-0.5i, l_{2}=0.5-0.5i, \alpha_{1}^{(1)}=0.65+0.5i, \alpha_{1}^{(2)}=1+i, \alpha_{2}^{(1)}=1+i, \alpha_{2}^{(2)}=0.65+0.5i, \alpha_{1}=\alpha_{2}=1, a=c=1$ and $b=0.5+0.5i$. These parameters are associated with the head-on collision between the two nondegenerate double-hump breathing solitons in both the components. As we can observe from the Figs. 6(a) and 6(b), before the collision the two individual nondegenerate double-hump breathing solitons are well separated in both the components and they preserve their shapes even after collision. One can visualize the shape preserving nature clearly from the corresponding contour plots depicted Figs. 6(c) and 6(d). 
\begin{figure}
	\centering
	\includegraphics[width=.35\linewidth]{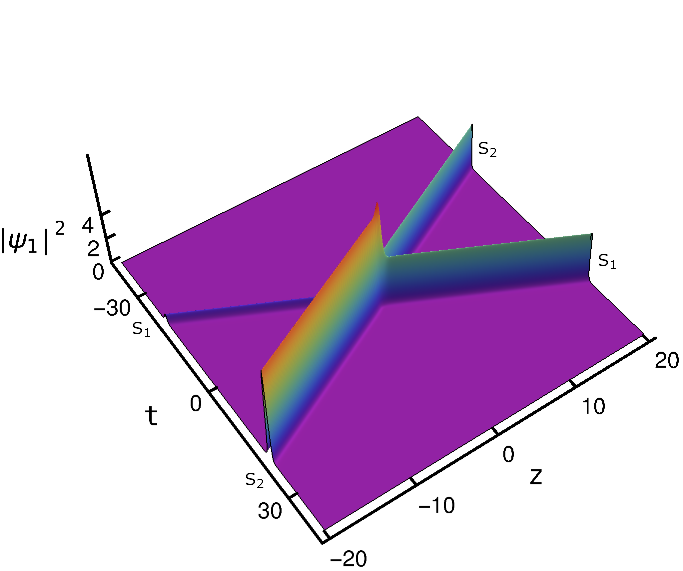}~
	\includegraphics[width=.35\linewidth]{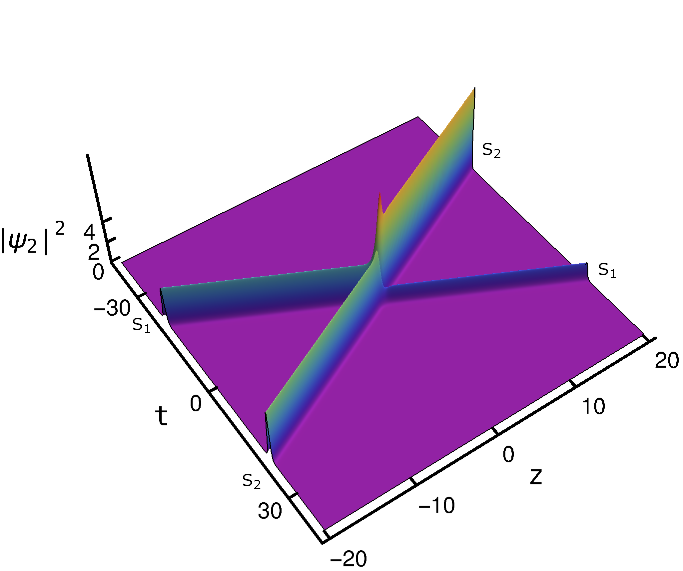}
	\caption{Shape changing collision between degenerate single-hump solitons. The parameter values: $k_{1}=l_{1}=1.5+0.5i, k_{2}=l_{2}=2-0.5i, \alpha_{1}^{(1)}=1.5, \alpha_{2}^{(1)}=0.45+0.9i, \alpha_{1}^{(2)}=\alpha_{2}^{(2)}=2+I, \alpha_{1}=\alpha_{2}=1, a=c=1$ and $b=0.5+0.5i$}
	\label{fig9}
\end{figure}
Next, we consider the situation of collision between degenerate and nondegenerate solitons. This can be studied by fixing the wave number restriction $k_{1}=l_{1}$ and $k_{2} \ne l_{2}$ in the two-soliton solution, given in Appendix A, of the GCNLS system (1). This coexistence of degenerate and nondegenerate solitons in both the components may be called the partially nondegenerate limit. This interesting collision of degenerate and nondegenerate solitons is shown in Fig. \ref{fig7} for the parameters $k_{1}=l_{1}=1.5+i, k_{2}=1.65-i, l_{2}=0.5-i, \alpha_{1}^{(1)}=\alpha_{2}^{(1)}=0.65, \alpha_{1}^{(2)}=\alpha_{2}^{(2)}=0.5i, \alpha_{1}=\alpha_{2}=1, a=c=1$ and $b=0.5+0.5i$. Similarly one can carry out the asymptotic analysis for this partially nondegenerate case as we described earlier for the complete nondegenerate case. In this case, there always occurs a finite energy redistribution among the coupled modes $\psi_{1}$ and $\psi_{2}$. In this scenario, as shown in Fig. \ref{fig7}, degenerate single-hump soliton interacts with the nondegenerate double-hump breathing soliton in both the components $\psi_{1}$ and $\psi_{2}$. After the collision, the intensity of the degenerate soliton gets reduced and the intensity of the nondegenerate double-hump breathing soliton is increased in the first component. In contrast, the intensity of the degenerate soliton is enhanced while the nondegenerate double-hump breathing soliton indensity is decreased in the second component. On the other hand this interesting energy redistribution between the two modes ensures the total energy conservation of the GCNLS system. One can also observe similar kind of shape changing collision among the nondegenerate and degenerate solitons for the wave number condition $k_{1} \neq l_{1}$ and $k_{2} = l_{2}$.

Pure degenerate two-soliton solutions of the system (1) can be deduced with the wave number restriction $k_{1} = l_{1}$ and $k_{2} = l_{2}$ in the two-soliton solution of (1). Under this condition, one can observe two single-hump solitons interact in both the components $\psi_{1}$ and $\psi_{2}$. Here, we have demonstrated the shape preserving collision between the degenerate solitons for the parameters $k_{1}=l_{1}=0.35+1.5i, k_{2}=l_{2}=0.45-1.5i, \alpha_{1}^{(1)}=\alpha_{2}^{(1)}=0.5+0.5i, \alpha_{1}^{(2)}=\alpha_{2}^{(2)}=0.45, \alpha_{1}=\alpha_{2}=1, a=c=1$ and $b=0.5+0.5i$. In Fig. \ref{fig8}, we can find the degenerate solitons preserve their shapes under collision. These degenerate solitons can also exhibit shape changing collision for other choice of parameters along with $\frac{\alpha_{1}^{(1)}}{\alpha_{2}^{(1)}} \neq \frac{\alpha_{1}^{(2)}}{\alpha_{2}^{(2)}}$. Shape changing collision of degenerate solitons is depicted in the Fig. \ref{fig9} with the parametric choices $k_{1}=l_{1}=1.5+0.5i, k_{2}=l_{2}=2-0.5i, \alpha_{1}^{(1)}=1.5, \alpha_{2}^{(1)}=0.45+0.9i, \alpha_{1}^{(2)}=\alpha_{2}^{(2)}=2+i, \alpha_{1}=\alpha_{2}=1, a=c=1$ and $b=0.5+0.5i$. In the $\psi_{1}$ component, the intensity of the degenerate single-hump soliton $S_{1}$ is getting increased after collision while the intensity of the single-hump soliton $S_{2}$ gets decreased. Exactly opposite kind of collision scenario occurs in the $\psi_{2}$ component where the intensity of the single-hump soliton $S_{1}$ is decreased under collision as the intensity of the single-hump soliton $S_{2}$ is increased. Hence, a finite energy redistribution occurrs among the two single-hump degenerate solitons that leads to shape changing interaction.

Finally, one can also investigate the collision scenerio in the case of nondegenerate bright-dark and dark-dark soliton cases associated with system (1). These are being pursued at present and they will be reported separately.

\section{Stability of linearly superimposed nondegenerate solitons}\label{sec4}
\begin{figure}
	\centering
	\includegraphics[width=.70\linewidth]{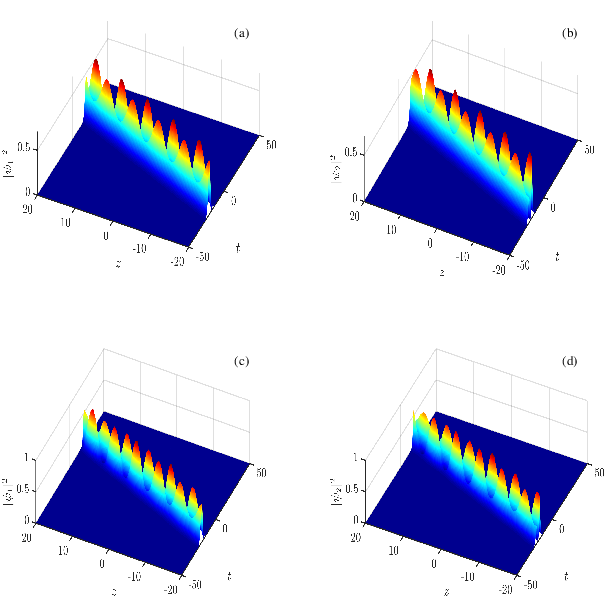}~~\\
	\includegraphics[width=.70\linewidth]{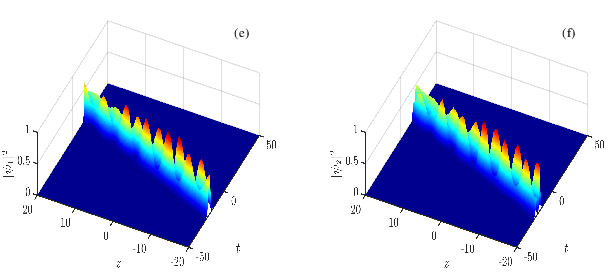}
	\caption{Numerical stability plots of nondegenerate double-hump breathing solitons (parameter values are same as given in Fig. (\ref{fig2})). (a) and (b): propagation of double-hump breathing solitons without adding any noise. (c) and (d): propagation of double-hump breathing solitons with 3$\%$ of Gaussian white noise. (e) and (f): propagation of double-hump breathing solitons with 5$\%$ of Gaussian white noise.}
	\label{fig10}
\end{figure}
In this section, we analyse the stability of the deduced linearly superimposed nondegenerate solitons of the GCNLS system (\ref{1}) numerically by adding various percentages of Gaussian white noise. We have performed the stability analysis for nondegenerate double-hump solitons, double-hump breathing solitons, and degenerate single-hump solitons against Gaussian white noise. Here we demonstrate the stability of the nondegenerate double-hump breathing solitons only.
To verify the analytical solutions of the GCNLS equation, direct numerical simulations were performed using a pseudo-spectral Split-Step Fourier Method (SSFM) with exact spectral integration of the linear operator and fourth-order Runge-Kutta (RK4) integration of the nonlinear operator.

The temporal computational domain was discretized uniformly over $t$ $\in$ $[t_{min}, t_{max}]$ using $N_t$ grid points, resulting in the temporal step size $\Delta t = \frac{t_{max} - t_{min}}{N_t}$, while the propagation interval $z$ $\in$ $[z_{min}, z_{max}]$ was divided into $N_z$ equal steps with $\Delta_z = \frac{z_{max} - z_{min}}{N_z}$. Unless otherwise stated, the simulations employed $N_t = 2^{10}$ (or $2^{13}$ for high-resolution
simulations) and $N_z = 5000$ (or 8000 for long-distance propagation), with the temporal and propagation step sizes determined from the corresponding computational windows.
The exact analytical solution derived in the Section \ref{sec2} was employed as the initial condition at $z = z_{min}$. To investigate the robustness of the analytical solutions against perturbations, complex random noise was added to the initial profile according to
\begin{equation}
	\psi_j(t, z_{min}) = \psi_j^{Analytical}(t, z_{min}) + \epsilon n_j(t),~~~ j = 1, 2.
\end{equation}
where $\epsilon$ denotes the prescribed perturbation amplitude (in percentage unit) and $n_j(t)$ represents independent zero-mean complex Gaussian random variables with unit variance. In the absence of perturbations, $\epsilon = 0$, and finite values of $\epsilon$ were used to examine the dynamical stability of the analytical solutions.
Since the Fast Fourier Transform inherently assumes periodic boundary conditions, the temporal computational window was chosen sufficiently large such that the field amplitudes at both boundaries remained close to machine precision throughout the propagation. Furthermore, a smooth absorbing layer was introduced near the temporal boundaries to suppress outgoing radiation and eliminate artificial wrap-around effects arising from the periodicity of the spectral method.

To perform the stability analysis, we have taken the initial profile from solution (\ref{4}) along with the initial conditions given in Fig. \ref{fig2}. We also fix the temporal domain $t$ as $[-50,50]$ and the propagation distance $z$ as $[-20,20]$. Through direct simulations we generate nondegenerate double-hump breathing solitons to system (\ref{1}) and propagate over distance $z$ without adding any additional noise. The same have been depicted in Figs. 10a-10b. Then to understand the stability against perturbation, we add 3$\%$ and 5$\%$ of Gaussian white noise with the above initial soliton solutions. Figures 10c-10d and 10e-10f correspond to the propagation of nondegenerate double-hump breathing solitons with the addition of 3$\%$ and 5$\%$ of Gaussian white noise, respectively. These figures confirm that the non-degenerate double-hump breathing soliton propagates in a stable manner in both components without any distortion and remains stable against 3$\%$ and 5$\%$ Gaussian white noise.

We also point out that, through a similar numerical analysis, the stability of the non-degenerate double-hump solitons and degenerate single-hump solitons was further examined under different levels of Gaussian white noise. The analysis revealed that all these solitons remain stable. However, the corresponding results are not presented here for brevity.

\section{Conclusion}\label{sec5}
In this paper, we have deduced the multi-nondegenerate soliton solutions of the GCNLS system (\ref{1}) from the nondegenerate bright soliton solutions of the Manakov-type system by using the linear interconnections between them. We found that such solutions exhibit breathing behaviour characterized by the corresponding frequency and period of oscillations. Then we have also systematically analyzed the role of the four-wave mixing parameter `$b$' and the linear coupling parameters $\alpha_{j}$, $j=1,2$. Through this analysis we realized that the linear coupling parameters play a major role in the breathing behavior of the deduced soliton solutions. We have captured the already known degenerate soliton solutions as a limiting case of the nondegenerate soliton solution by imposing the conditions on the wave numbers appropriately. Further, we have carried out the systematic asymptotic analysis and explicitly shown the relation between the phases of the superimposed nondegenerate solitons before and after the collision process. From this analysis we have explained the shape-preserving collision of superimposed nondegenerate double-hump breathing solitons. We have also demonstrated the interesting shape-changing collision nature between the degenerate and double-hump breathing nondegenerate soliton under partially nondegenerate limit of the superimposed two-soliton solution. In the purely degenerate limit, we have demonstrated the shape-preserving collision and interesting shape-changing collisions due to the energy redistribution among the linearly transformed degenerate solitons. Also, we have numerically analyzed the stability of the deduced nondegenerate double-hump breathing solitons against different percentages of Gaussian white noise. The results presented in this paper will be useful for understanding the linearly transformed vector solitons in the soliton supporting coupled nonlinear Schr\"{o}dinger family of equations and related field of research in soliton theory.
\backmatter

\bmhead{Acknowledgements}

The authors are thankful to Dr. A. Govindarajan, Visiting Scientist, Department of Nonlinear Dynamics, Bharathidasan University, Tiruchirappalli for his help in verifying the numerical stability.
R.R. would like to acknowledge the financial support in the form of DST-ANRF National Post-doctoral Fellowship (File No. PDF/2023/001115).
M.L. thanks DST-ANRF, INDIA for the award of a DST-ANRF National Science Chair (NSC/2020/000029) position in which S.S. is a Research Associate.

\bmhead{Data Availability Statements}

No new data were created or analysed in this study.

\bmhead{Conflict of Interest}

The authors declare that they have no conflict of interest.
\begin{appendices}
	\section{Nondegenerate one- and two-soliton solutions of the Manakov-type system (\ref{3})}
	The nondegenerate one-soliton solution of the Manakov-type system given in (\ref{3}) in the Gram determinant form as reported in ref.\cite{rkpre} reads as
	\begin{equation*} 
		q_{1}=\frac{g^{(1)}}{f},~ q_{2}=\frac{g^{(2)}}{f},
	\end{equation*}
	%
	where
	\begin{eqnarray*}
		g^{(1)} =
		\begin{vmatrix}
			\frac{e^{\eta_1+\eta_1^*}}{(k_1+k_1^*)} & \frac{e^{\eta_1+\xi_1^*}}{(k_1+l_1^*)} & 1 & 0 & e^{\eta_1} \\ 
			\frac{e^{\xi_1+\eta_1^*}}{(l_1+k_1^*)} & \frac{e^{\xi_1+\xi_1^*}}{(l_1+l_1^*)}  & 0 & 1 &   e^{\xi_1}\\
			-1 & 0 & \frac{|\al_1^{(1)}|^2 \sigma_{1}}{(k_1+k_1^*)} & 0 & 0\\
			0 & -1 & 0 &  \frac{|\al_1^{(2)}|^2 \sigma_{2}}{(l_1+l_1^*)} & 0\\
			0 & 0& -\al_1^{(1)} & 0 &0
		\end{vmatrix},~~
		g^{(2)} =
		\begin{vmatrix}
			\frac{e^{\eta_1+\eta_1^*}}{(k_1+k_1^*)} & \frac{e^{\eta_1+\xi_1^*}}{(k_1+l_1^*)} & 1 & 0 & e^{\eta_1} \\ 
			\frac{e^{\xi_1+\eta_1^*}}{(l_1+k_1^*)} & \frac{e^{\xi_1+\xi_1^*}}{(l_1+l_1^*)}  & 0 & 1 &   e^{\xi_1}\\
			-1 & 0 & \frac{|\al_1^{(1)}|^2 \sigma_{1}}{(k_1+k_1^*)} & 0 & 0\\
			0 & -1 & 0 &  \frac{|\al_1^{(2)}|^2 \sigma_{2}}{(l_1+l_1^*)} & 0\\
			0 & 0& 0 & -\al_1^{(2)} &0
		\end{vmatrix},~~
	\end{eqnarray*}
	and
	\begin{eqnarray*}
		f=
		\begin{vmatrix}
			\frac{e^{\eta_1+\eta_1^*}}{(k_1+k_1^*)} & \frac{e^{\eta_1+\xi_1^*}}{(k_1+l_1^*)} & 1 & 0  \\ 
			\frac{e^{\xi_1+\eta_1^*}}{(l_1+k_1^*)} & \frac{e^{\xi_1+\xi_1^*}}{(l_1+l_1^*)}  & 0 & 1 \\
			-1 & 0 & \frac{|\al_1^{(1)}|^2 \sigma_{1}}{(k_1+k_1^*)} & 0 \\
			0 & -1 & 0 &  \frac{|\al_1^{(2)}|^2 \sigma_{2}}{(l_1+l_1^*)}
		\end{vmatrix}.
	\end{eqnarray*}
	Similarly, the Gram determinants of $g^{(1)}$, $g^{(2)}$ and $f$ for the nondegenerate two-soliton solution of the Manakov-type system (\ref{3}) have the following forms:
	\begin{eqnarray*}
		g^{(1)} =
		\begin{vmatrix}
			\frac{e^{\eta_1+\eta_1^*}}{(k_1+k_1^*)} & \frac{e^{\eta_1+\eta_2^*}}{(k_1+k_2^*)} & \frac{e^{\eta_1+\xi_1^*}}{(k_1+l_1^*)} & \frac{e^{\eta_1+\xi_2^*}}{(k_1+l_2^*)} & 1 & 0 & 0& 0 & e^{\eta_1} \\ 
			\frac{e^{\eta_2+\eta_1^*}}{(k_2+k_1^*)} & \frac{e^{\eta_2+\eta_2^*}}{(k_2+k_2^*)} & \frac{e^{\eta_2+\xi_1^*}}{(k_2+l_1^*)} & \frac{e^{\eta_2+\xi_2^*}}{(k_2+l_2^*)} & 0 & 1 & 0& 0 & e^{\eta_2} \\ 
			\frac{e^{\xi_1+\eta_1^*}}{(l_1+k_1^*)} & \frac{e^{\xi_1+\eta_2^*}}{(l_1+k_2^*)} & \frac{e^{\xi_1+\xi_1^*}}{(l_1+l_1^*)} & \frac{e^{\xi_1+\xi_2^*}}{(l_1+l_2^*)} & 0 & 0 & 1& 0 & e^{\xi_1} \\ 
			\frac{e^{\xi_2+\eta_1^*}}{(l_2+k_1^*)} & \frac{e^{\xi_2+\eta_2^*}}{(l_2+k_2^*)} & \frac{e^{\xi_2+\xi_1^*}}{(l_2+l_1^*)} & \frac{e^{\xi_2+\xi_2^*}}{(l_2+l_2^*)} & 0 & 0 & 0& 1 & e^{\xi_2} \\ 
			-1 & 0 & 0& 0& \frac{|\al_1^{(1)}|^2\sigma_1}{(k_1^*+k_1)} & \frac{\al_1^{(1)*}\al_2^{(1)}\sigma_1}{(k_1^*+k_2)} & 0 &0 & 0\\
			0 & -1 & 0 &  0& \frac{\al_1^{(1)}\al_2^{(1)*}\sigma_1}{(k_2^*+k_1)}& \frac{|\al_2^{(1)}|^2\sigma_1}{(k_2+k_2^*)} & 0 & 0 & 0\\
			0 & 0 & -1 &  0& 0& 0 & \frac{|\al_1^{(2)}|^2\sigma_2}{(l_1^*+l_1)} & \frac{\al_1^{(2)*}\al_2^{(2)}\sigma_2}{(l_1^*+l_2)} & 0\\
			0 & 0 & 0 & -1& 0& 0& \frac{\al_1^{(2)}\al_2^{(2)*}\sigma_2}{(l_2^*+l_1)} & \frac{|\al_2^{(2)}|^2\sigma_2}{(l_2^*+l_2)} & 0\\
			0 & 0& 0& 0& -\al_1^{(1)} & -\al_2^{(1)} &  0 &0  &0
		\end{vmatrix},\label{3.7a}
	\end{eqnarray*}
	\begin{eqnarray*}
		g^{(2)} =
		\begin{vmatrix}
			\frac{e^{\eta_1+\eta_1^*}}{(k_1+k_1^*)} & \frac{e^{\eta_1+\eta_2^*}}{(k_1+k_2^*)} & \frac{e^{\eta_1+\xi_1^*}}{(k_1+l_1^*)} & \frac{e^{\eta_1+\xi_2^*}}{(k_1+l_2^*)} & 1 & 0 & 0& 0 & e^{\eta_1} \\ 
			\frac{e^{\eta_2+\eta_1^*}}{(k_2+k_1^*)} & \frac{e^{\eta_2+\eta_2^*}}{(k_2+k_2^*)} & \frac{e^{\eta_2+\xi_1^*}}{(k_2+l_1^*)} & \frac{e^{\eta_2+\xi_2^*}}{(k_2+l_2^*)} & 0 & 1 & 0& 0 & e^{\eta_2} \\ 
			\frac{e^{\xi_1+\eta_1^*}}{(l_1+k_1^*)} & \frac{e^{\xi_1+\eta_2^*}}{(l_1+k_2^*)} & \frac{e^{\xi_1+\xi_1^*}}{(l_1+l_1^*)} & \frac{e^{\xi_1+\xi_2^*}}{(l_1+l_2^*)} & 0 & 0 & 1& 0 & e^{\xi_1} \\ 
			\frac{e^{\xi_2+\eta_1^*}}{(l_2+k_1^*)} & \frac{e^{\xi_2+\eta_2^*}}{(l_2+k_2^*)} & \frac{e^{\xi_2+\xi_1^*}}{(l_2+l_1^*)} & \frac{e^{\xi_2+\xi_2^*}}{(l_2+l_2^*)} & 0 & 0 & 0& 1 & e^{\xi_2} \\ 
			-1 & 0 & 0& 0& \frac{|\al_1^{(1)}|^2\sigma_1}{(k_1^*+k_1)} & \frac{\al_1^{(1)*}\al_2^{(1)}\sigma_1}{(k_1^*+k_2)} & 0 &0 & 0\\
			0 & -1 & 0 &  0& \frac{\al_1^{(1)}\al_2^{(1)*}\sigma_1}{(k_2^*+k_1)}& \frac{|\al_2^{(1)}|^2\sigma_1}{(k_2+k_2^*)} & 0 & 0 & 0\\
			0 & 0 & -1 &  0& 0& 0 & \frac{|\al_1^{(2)}|^2\sigma_2}{(l_1^*+l_1)} & \frac{\al_1^{(2)*}\al_2^{(2)}\sigma_2}{(l_1^*+l_2)} & 0\\
			0 & 0 & 0 & -1& 0& 0& \frac{\al_1^{(2)}\al_2^{(2)*}\sigma_2}{(l_2^*+l_1)} & \frac{|\al_2^{(2)}|^2\sigma_2}{(l_2^*+l_2)} & 0\\
			0 & 0& 0& 0& 0& 0& -\al_1^{(2)} & -\al_2^{(2)}  &0
		\end{vmatrix},\label{3.7b}
	\end{eqnarray*}
	and
	\begin{eqnarray*}
		f=
		\begin{vmatrix}
			\frac{e^{\eta_1+\eta_1^*}}{(k_1+k_1^*)} & \frac{e^{\eta_1+\eta_2^*}}{(k_1+k_2^*)} & \frac{e^{\eta_1+\xi_1^*}}{(k_1+l_1^*)} & \frac{e^{\eta_1+\xi_2^*}}{(k_1+l_2^*)} & 1 & 0 & 0& 0 \\ 
			\frac{e^{\eta_2+\eta_1^*}}{(k_2+k_1^*)} & \frac{e^{\eta_2+\eta_2^*}}{(k_2+k_2^*)} & \frac{e^{\eta_2+\xi_1^*}}{(k_2+l_1^*)} & \frac{e^{\eta_2+\xi_2^*}}{(k_2+l_2^*)} & 0 & 1 & 0& 0 \\ 
			\frac{e^{\xi_1+\eta_1^*}}{(l_1+k_1^*)} & \frac{e^{\xi_1+\eta_2^*}}{(l_1+k_2^*)} & \frac{e^{\xi_1+\xi_1^*}}{(l_1+l_1^*)} & \frac{e^{\xi_1+\xi_2^*}}{(l_1+l_2^*)} & 0 & 0 & 1& 0\\ 
			\frac{e^{\xi_2+\eta_1^*}}{(l_2+k_1^*)} & \frac{e^{\xi_2+\eta_2^*}}{(l_2+k_2^*)} & \frac{e^{\xi_2+\xi_1^*}}{(l_2+l_1^*)} & \frac{e^{\xi_2+\xi_2^*}}{(l_2+l_2^*)} & 0 & 0 & 0& 1 \\ 
			-1 & 0 & 0& 0& \frac{|\al_1^{(1)}|^2\sigma_1}{(k_1^*+k_1)} & \frac{\al_1^{(1)*}\al_2^{(1)}\sigma_1}{(k_1^*+k_2)} & 0 &0\\
			0 & -1 & 0 &  0& \frac{\al_1^{(1)}\al_2^{(1)*}\sigma_1}{(k_2^*+k_1)}& \frac{|\al_2^{(1)}|^2\sigma_1}{(k_2+k_2^*)} & 0 & 0\\
			0 & 0 & -1 &  0& 0& 0 & \frac{|\al_1^{(2)}|^2\sigma_2}{(l_1^*+l_1)} & \frac{\al_1^{(2)*}\al_2^{(2)}\sigma_2}{(l_1^*+l_2)}\\
			0 & 0 & 0 & -1& 0& 0& \frac{\al_1^{(2)}\al_2^{(2)*}\sigma_2}{(l_2^*+l_1)} & \frac{|\al_2^{(2)}|^2\sigma_2}{(l_2^*+l_2)}
		\end{vmatrix}.\label{3.7c}
	\end{eqnarray*}
\end{appendices}
\end{document}